\documentclass{PoS}

\usepackage{amsfonts,amssymb,amsmath,bm}
\usepackage{graphicx}

\title{Gluon Mass, Glueballs and Gluonic Mesons}

\ShortTitle{Gluon Mass, Glueballs and Gluonic Mesons}

\author{\speaker{Vincent Mathieu}\thanks{CPAN postdoc}\\
        Departament de F\'{\i}sica Te\`orica and Institut de F\'{\i}sica Corpuscular,\\
Universitat de Val\`encia-CSIC, E-46100 Burjassot (Valencia), Spain.\\
        E-mail: \email{vincent.mathieu@umons.ac.be}}

\abstract{We review the phenomenological and theoretical evidences for dynamical gluon mass generation and the main features of the glueball spectrum in (pure gauge) Yang-Mills theories. The mixing between glueball and conventional $\bar q q$ states in $f_0$ scalar mesons is discussed. For pseudoscalar mesons, the inclusion of the glue field in an effective low energy theory is presented leading to a third isoscalar $\eta''$ partner of the $\eta$ and $\eta'$. Branching ratios for processes involving $\eta''$ are given and, when available, compatible with data for $\eta(1405)$.}

\FullConference{The many faces of QCD\\
		November 2-5, 2010\\
		Gent Belgium}

\begin{document}

\section{Introduction}
The gluon is the fundamental particle of Yang-Mills theories. Gauge invariance forbids a mass term in the Lagrangian. The gluon is therefore a massless particle. However it has been argued long time ago~\cite{Cornwall:1981zr} that non-perturbative effects might lead to a dynamical mass for the gluon without breaking gauge invariance. This dynamically generated mass appeared to be the cure for the infrared slavery of non-Abelian gauge theories~\cite{PTbook,Aguilar:2008xm}.

In section 2, we review the implications of the gluon mass generation. The reasons why we do not observe a longitudinal component for the gluon is summarized in section 3. Section 4 concerns the spectrum of pure gauge quantum chromodynamics (QCD). We next turn our attention to mixing between glueballs and $\bar q q$ states in scalar mesons in section 5. The inclusion of the glueball in the chiral Lagrangien is presented in section 6 where predictions are given for the third partner associated to $\eta$ and $\eta'$. Section 7 summarizes the conclusions.

\section{Evidences for dynamical mass generation}

We have several reasons to believe that dynamical mass generation occurs in QCD.

If we believe in QCD to properly describe in the strong interaction, we should ask for its consistency in the infrared where the theory become strongly coupled. The negative sign of the $\beta-$function first coefficient $b$ clearly implies imaginary poles, associated with non physical particles, for small $q^2$ in all order in perturbation theory (infrared slavery). The non perturbative generation of a gluon mass solves this problem and renders the theory consistent in the deep infrared, see the discussion in section 2.2.3 of ref~\cite{PTbook}.

\begin{figure}[htb]
        \centerline{\includegraphics[width=0.7\linewidth]{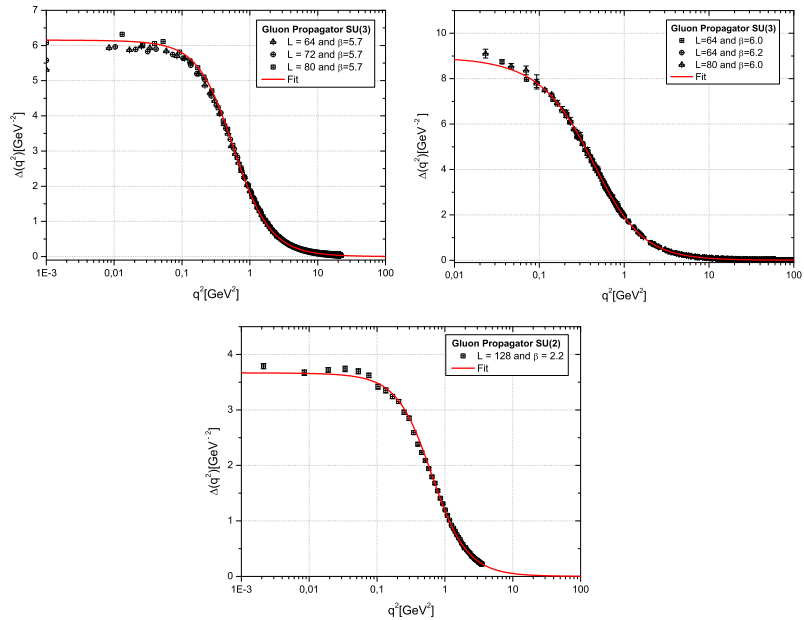}}
        \caption{\label{fig:lattprop} Gluon propagator in the Landau gauge. Data from \cite{Cucchieri:2007md,Bogolubsky:2007ud,Oliveira:2009nn}.}
\end{figure}

Recently accurate lattice studies showed an infrared saturation of the gluon propagator in the Landau gauge, see Fig.~\ref{fig:lattprop}. Those data are compatible with a massive-like propagator, solution of the corresponding Dyson-Schwinger equation~\cite{Aguilar:2008xm,Cornwall:1981zr,Binosi:2009qm}
\begin{equation}\label{eq:gluon_prop_cornwall}
d^{-1}(q^2) = \left(q^2+m^2(q^2)\right) bg^2 \ln\left[\frac{q^2+4m^2(q^2)}{\Lambda^2}\right],
\end{equation}
with a dynamical running mass
\begin{equation}\label{eq:dym_mass}
    m^2(q^2) =
    m_0^2\left(\frac{\ln\left[(q^2+4m_0^2)/\Lambda^2\right]}
    {\ln\left(4m_0^2/\Lambda^2\right)}\right)^{-12/11}.
\end{equation}
Exactly as happens with the quark constituent mass,  the effective gluon mass depends non-trivially on the momentum, vanishing in the deep ultraviolet.

\begin{figure}[htb]
        \centerline{\includegraphics[width=0.435\linewidth]{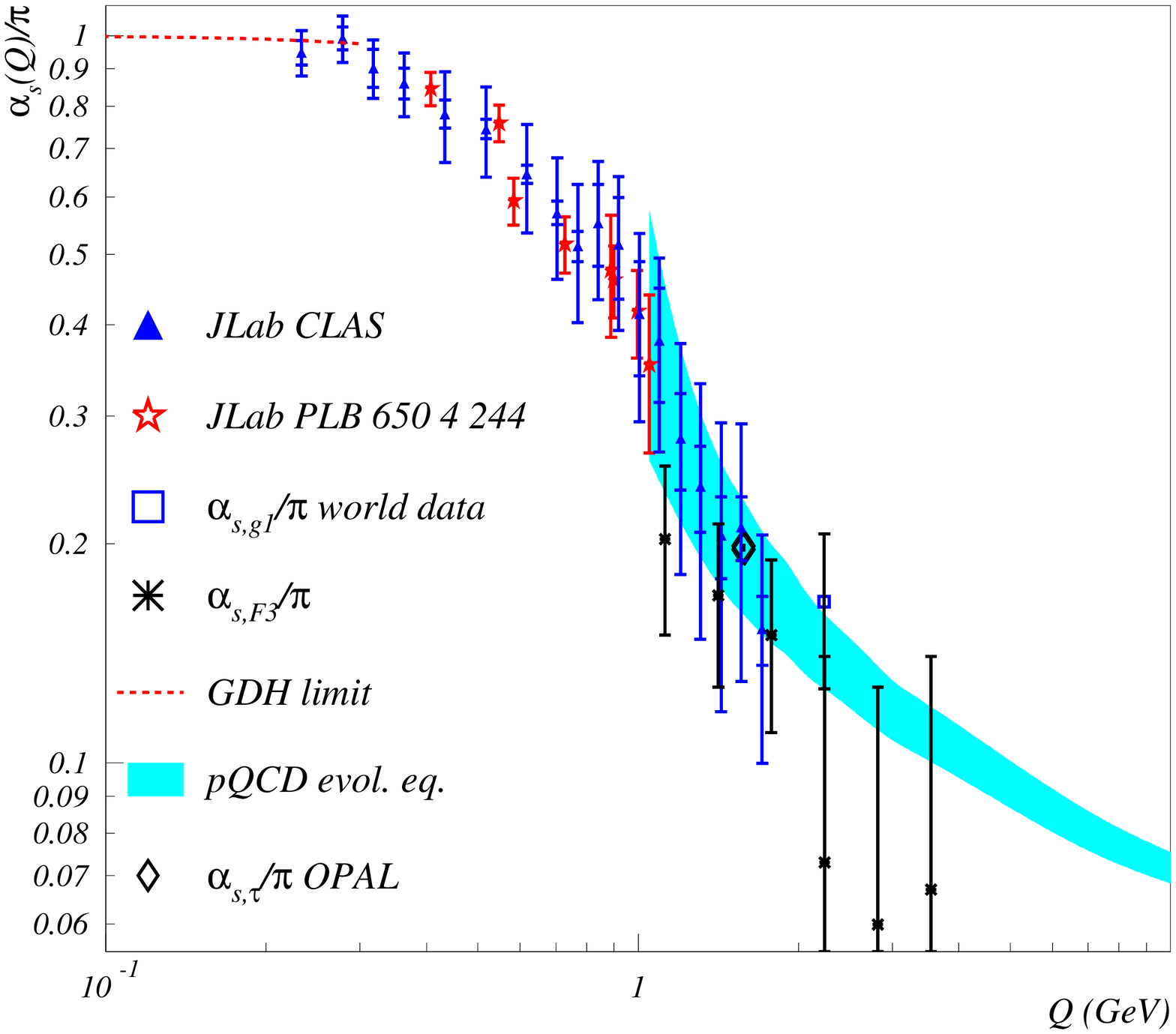}
        \includegraphics[width=0.35\linewidth]{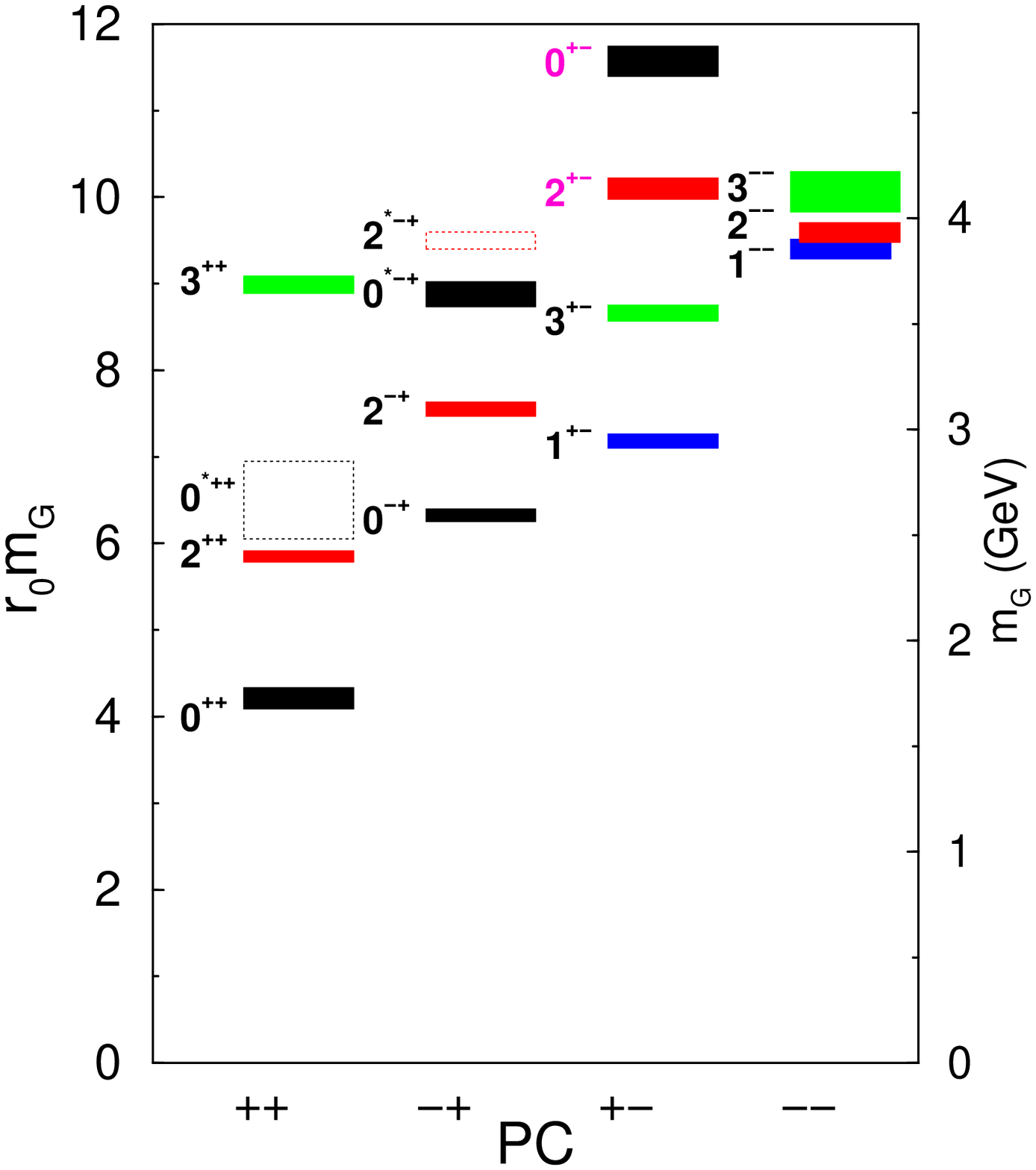}}
        \caption{\label{fig:alphas} {\it Right, }strong coupling constant at large distances, Extracted form Ref.~\cite{Deur:2009tj}; {\it left,} continuum glueball spectrum from lattice QCD~\cite{Morningstar:1999rf}.}
\end{figure}

Massive solutions lead naturally to the concept of non Abelian effective charge freezing in the infrared~\cite{Binosi:2009qm}. Data from various processes at small $q^2$ indeed shows hints for such a behaviour, see Fig.~\ref{fig:alphas}. This saturation of the strong coupling constant is also a justification of hadronic models used in perturbation treatments of transverse momentum dependent parton distribution functions~\cite{Courtoy:2011mf}. Moreover, non perturbative power corrections to event shapes (such as the Trust), reveals a common value for the infrared averaged strong coupling constant fitted from data~\cite{Dokshitzer:1995zt},
\begin{equation}\label{}
\mu_I^{-1}\int_0^{\mu_I}\alpha(Q^2)dQ^2\sim 0.5,
\end{equation}
in disagreement with a would-be power law scaling of the strong coupling constant.

The (OZI forbidden) $J/\psi$ radiative decays is the typical example of glue rich processes which probes gluon propagators in the infrared. Manifestations of gluon mass was investigated in the photon inclusive $J/\psi\to\gamma X$ decays with a important improvement of the shape with the inclusion of a gluon mass~\cite{Field:2001iu}. Data favors a value $m_g\sim 700$ MeV in agreement with other models (for a review of the concept of gluon mass in various model see~\cite{Mathieu:2008me}).

Finally, the existence of bound states in pure Yang-Mills theories, glueballs~\cite{Mathieu:2008me}, tell us that gluon gains an effective mass due to confinement in the same way that quarks have an effective running mass in mesons and baryons. We have no reason to believe that gluon mass generation will not be also realized in QCD in the presence of sea quarks.

\section{Only two degrees of freedom}
The evidences summarized in the previous sections reflect the consequences of mass generation but not the causes. Our intellect requires a understanding of this mechanism at a fundamental level. What makes the gluon massive is a non-Abelian version of the well-known Schwinger mechanism~\cite{Schwinger:1962tn}. Mass generation without spoiling gauge invariance and without scalar fields were investigated already back to the 70's~\cite{Jackiw:1973tr}. It is shown that a simple ansatz for the tree-gluon vertex leads consistently to a mass in the propagator respecting the Slavnov-Taylor identity. This ansatz for the vertex can be guessed form a massive gauge invariant QCD, where in addition to the conventional Yang-Mills Lagrangian a non linear gauge sigma model is supplied providing the appropriate extra degrees of freedom~\cite{Cornwall:1974hz}.

Because, indeed, a massive vector particle requires a third longitudinal component with respect to a massless one. We then arrive to the question about the physical relevance of the third component. As it can be shown~\cite{Jackiw:1973tr}, this scalar pole triggering mass generation through the so-called Schwinger mechanism, decouples from on-shell $S-$matrix due to current conservation. Moreover, as explained in details in~\cite{Mathieu:2009cc}, this third component do not appear neither in glueball wave functions.

\section{Glueballs in Yang-Mills theories}
The spectrum of pure gauge theories was investigated from various points of view~\cite{Mathieu:2008me}. The spectrum of low-lying glueballs obtained by Morningstar and Peardon is presented in Fig.~\ref{fig:alphas}. They restricted their study to low dimensional gluonic operators and states below four GeV. Although they did not draw any definitive conclusion concerning a $1^{++}$ state, they found a clear signal for a vector state but above the two-glueball molecule threshold~\cite{Morningstar:1999rf}.

It has been argue that no two-glueball vector state exists in agreement with Yang's theorem. This idea deserves clarification. A vector do exist for non Abelian gauge group and appear in the decomposition of the tensor built out of two gluon field strength
$G^a_{\mu\nu}D_\delta G^a_{\alpha\beta}$~\cite{Jaffe:1985qp}.
This is not in contradiction with Yang's theorem saying that a vector meson cannot decay into two massless vector particles. One has just to keep in mind that (the non Abelian part of) $G^a_{\mu\nu}$ involves more than one gluon operators. At the level of constituent model, it is not possible to construct of vector wave function out of two transverse gluons~\cite{Mathieu:2009cc} and indeed the vector signal found in the lattice study~\cite{Morningstar:1999rf} is a mass gap above the two-gluon glueballs.

Constituent models teach us that only two gluonic degrees of freedom are required by each gluon in the wave function to reproduce properly the lattice spectrum~\cite{Mathieu:2009cc}. We learn also from this technique that instanton contributions play an important role in scalar and pseudoscalar correlators. This is supported by Forkel's analysis using QCD spectral sum rules~\cite{Forkel:2003mk}. Another support for the massive gluon propagator and the instanton importance for scalar operators comes by the recent analysis of Dudal {\it et al.}~\cite{Dudal:2010cd}. Using a massive gluon propagators fitted from lattice calculations, they computed the one loop gluonic operators for the lowest states (no instantons contributions included). After some subtractions, they found masses in perfect agreement with constituent models
\begin{equation}\label{}
    M_{0^{++}} = 1.96\text{ GeV,} \qquad M_{0^{-+}} = 2.19\text{ GeV}.
\end{equation}
Moreover, this is a clear indication that the fully dressed propagator can play the role of the condensates in the operator product expansion.

\section{Scalar mesons}
\begin{figure}[htb]
       \centerline{\includegraphics[width=1.2\linewidth]{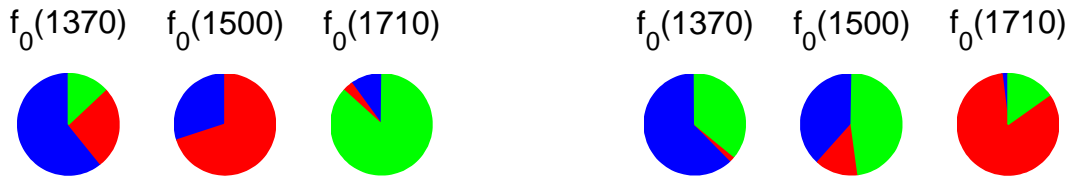}}
       \caption{\label{fig:mixingscalar}Mixing schemes for isoscalar mesons from Cheng {\it et al.}~\cite{Cheng:2006hu} (left) and from Close and Kirk~\cite{Close:2000yk} (right) ; blue: $\bar q q$ ; red: $\bar s s$ ; green: $gg$}
\end{figure}

Three isoscalar would have been observed in central production~\cite{Crede:2008vw}:
\begin{equation}\label{}
    f_0(1370) \qquad f_0(1500) \qquad f_0(1710)
\end{equation}
Although no definitive conclusion about their existence can be drawn~\cite{Crede:2008vw}, three isoscalar would imply a mixing between the two conventional $\bar q q$ and $\bar s s$ with a glueball ($gg$). We call mesons with a large glue content, {\it gluonic mesons}. With the discovery of the $f_0(1500)$, Close interpreted it as a glueball candidate and predict a third isoscalar gluonic meson to be discovered later on. With the discovery of the $f_0(1710)$ coupling stronger to $K\bar K$ than to $\pi\pi$, Close and Kirk proposed a mixing scheme, Fig.~\ref{fig:mixingscalar} (right), where the glueball is shared between the three isoscalar~\cite{Close:2000yk}. In this interpretation, the heaviest state is mainly a $\bar s s$ meson due to its coupling to $K \bar K$.

However, Chanowitz showed that the scalar glueball couples to $\bar q q$ with a strength proportional to the quark mass~\cite{Chanowitz:2005du}. Using this chiral suppression argument and lattice inputs for the bare masses, Cheng {\it et al} proposed another scheme, Fig.~\ref{fig:mixingscalar} (left), where the $f_0(1710)$ is mainly the glueball~\cite{Cheng:2006hu}.

The situation is even more obscur in view of B factories (Belle and Babar) results~\cite{Crede:2008vw}: The invariant $K^+K^-$ mass shows a peak around 1500 MeV/c2 denoted with mass and width consistent with the standard f0(1500) state. An observation
of $f0(1500)\to K^+K^-$, but no signal in the decay to $\pi^+\pi^-$ is inconsistent with the standard f0(1500), which is expected to couple more strongly to the two-pion decay.

\section{Pseudoscalar mesons}
The first pseudoscalar glueball candidate was observed in $J/\psi$ radiative decays by the Mark III collaboration~\cite{Crede:2008vw}. Actually, they observed two resonances denoted $\eta(1405)$ and $\eta(1475)$. Only the latter was observed in $\gamma\gamma$ fusion, leading to a possible large glue content in the former. The actual interpretation favors $\eta$ and $\eta'$ radial excitations for $\eta(1295)$ and $\eta(1475)$ and leave $\eta(1405)$ as the glueball candidate.

In addition to a pseudoscalar gluonic meson candidate, we have indication of a possible large glue content in the $\eta'$ wave function. $J/\psi$ radiative decay is a gluon rich environnement and the experimental branching ratio shows a large coupling for the $\eta'$
\begin{equation*}
    \frac{\Gamma(J/\psi\to\eta^\prime\gamma)}{\Gamma(J/\psi\to\eta \gamma)} = \left(\frac{\langle0|G\tilde G|\eta'\rangle}{\langle0|G\tilde G|\eta\rangle}\right)^2 \left(\frac{M_{J/\psi}^2-M_{\eta^\prime}^2}{M_{J/\psi}^2-M_{\eta}^2}\right)^3
    =4.81 \pm 0.77.
\end{equation*}

It could be therefore interesting to have a theoretical framework to study the mixing between the group theoretical states $\eta_0$ and $\eta_8$, and the pseudoscalar glueball $\eta_g$.
The chiral Lagrangian in the large-$N$ provides such a tools. The singlet $\eta_0$ is included in the non linear parametrization for the Goldstone bosons $U=\exp\left(i\sqrt{2}\pi/f\right)$ with $\pi = \pi^a\lambda_a$ ($\lambda_0\equiv\bm1_3/\sqrt{3}$) and, at each order in $p^2$, only the leading term in $N$ is kept.

In order to investigate the mixing with glue, one has to couple it to Goldstone bosons. Such a coupling is provided by the anomaly since the anomalous operator $\tilde G_{\mu\nu} G^{\mu\nu}$ interpolates the pseudoscalar glueball. At the effective level, we add a kinetic term and a mass term for $\eta_g$ coupled to $\eta_0$ {\it via} the anomaly~\cite{Rosenzweig:1981cu,Mathieu:2009sg} and we obtain at leading order
\begin{equation}
    {\cal L}^{(p^2)} = \frac{f^2}{8}\left<\partial_\mu U^\dag\partial^\mu U + B( m U^\dag  + U m^\dag)\right>  - \frac{\alpha}{2}(\eta_0+ k\eta_g)^2 - \frac{1}{2}m_\theta^2 \eta_g^2 + \frac{1}{2}\partial_\mu \eta_g\partial^\mu \eta_g.
\end{equation}

\begin{figure}[htb]
        \includegraphics[width=0.33\linewidth]{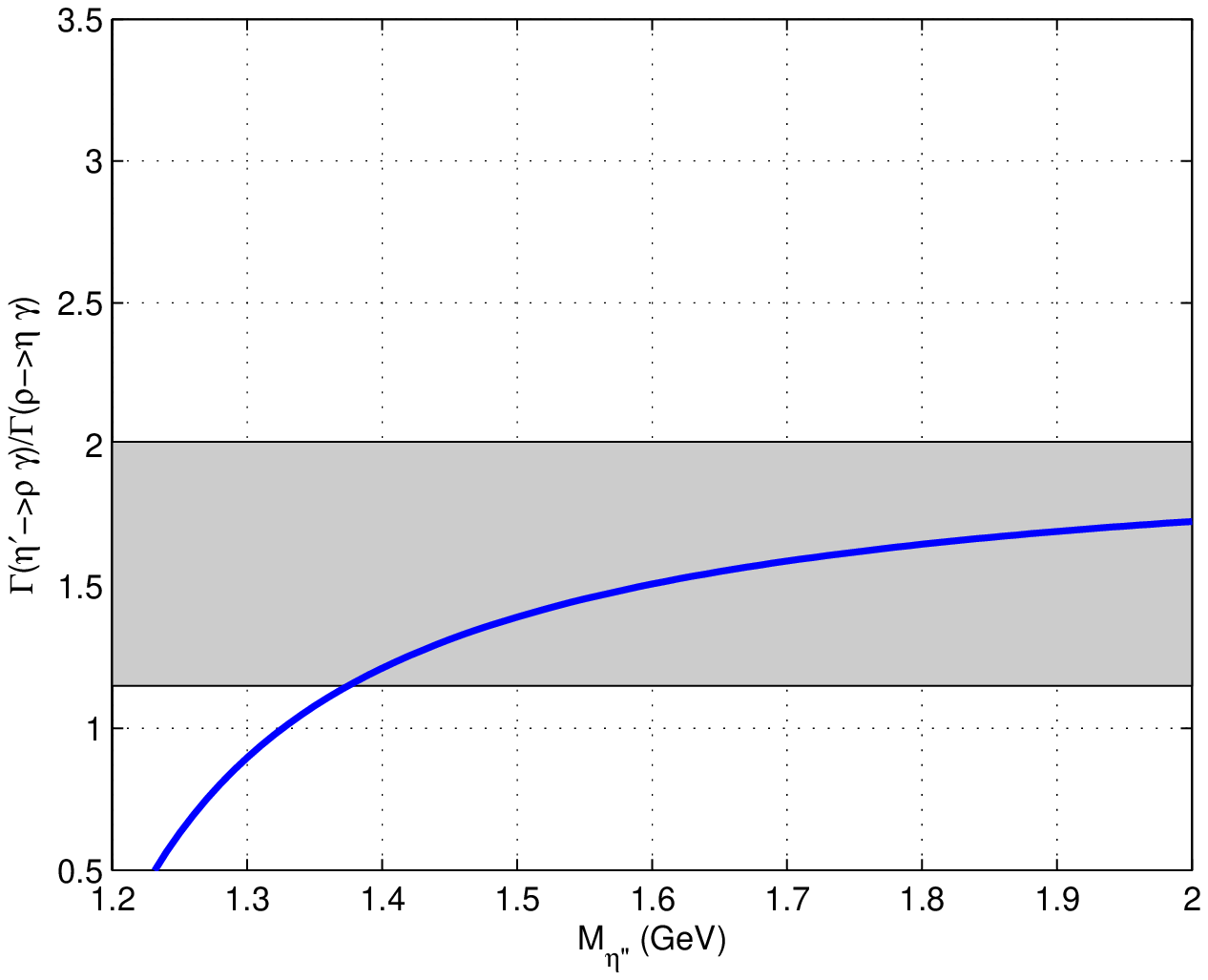}
        \includegraphics[width=0.33\linewidth]{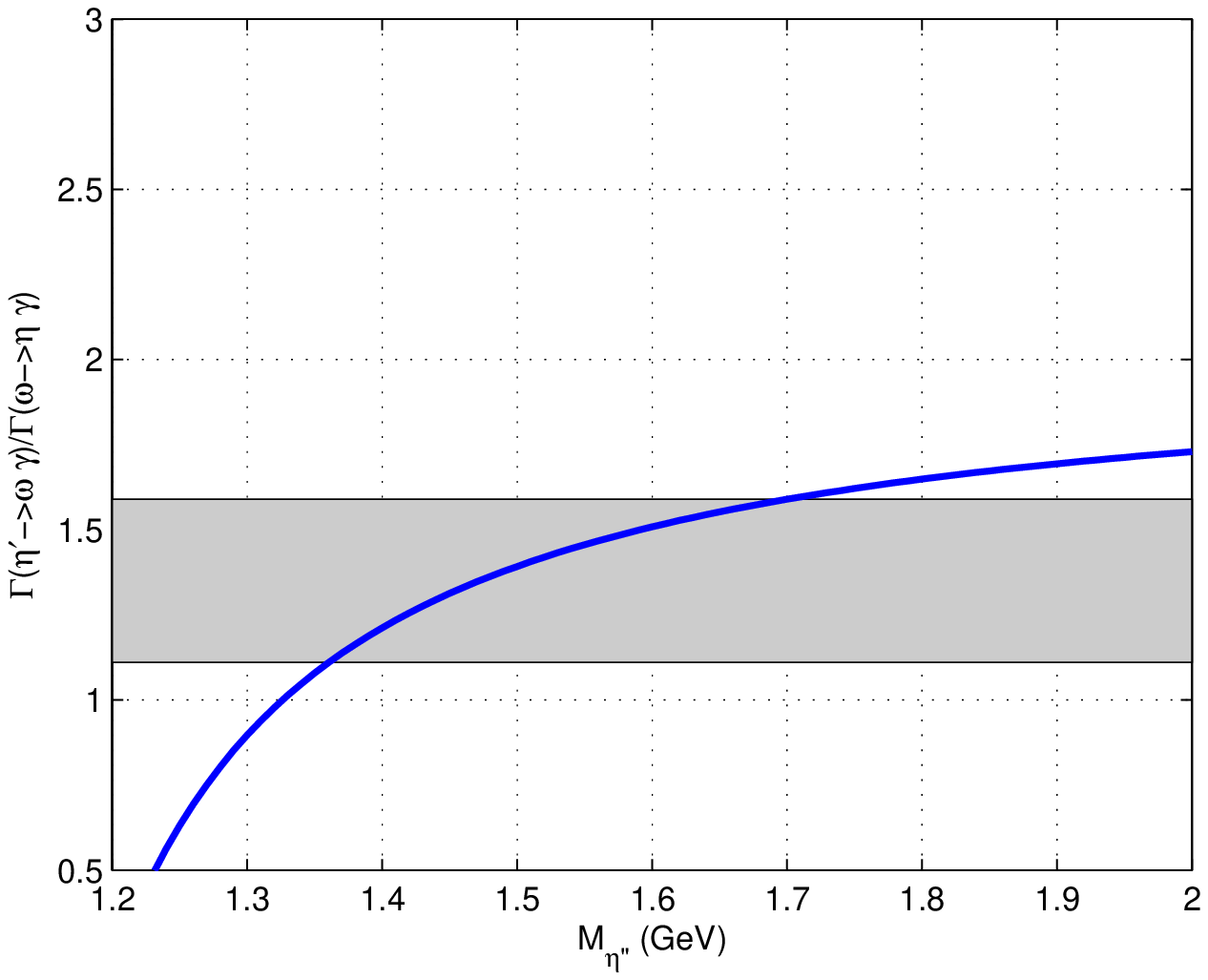}
        \includegraphics[width=0.33\linewidth]{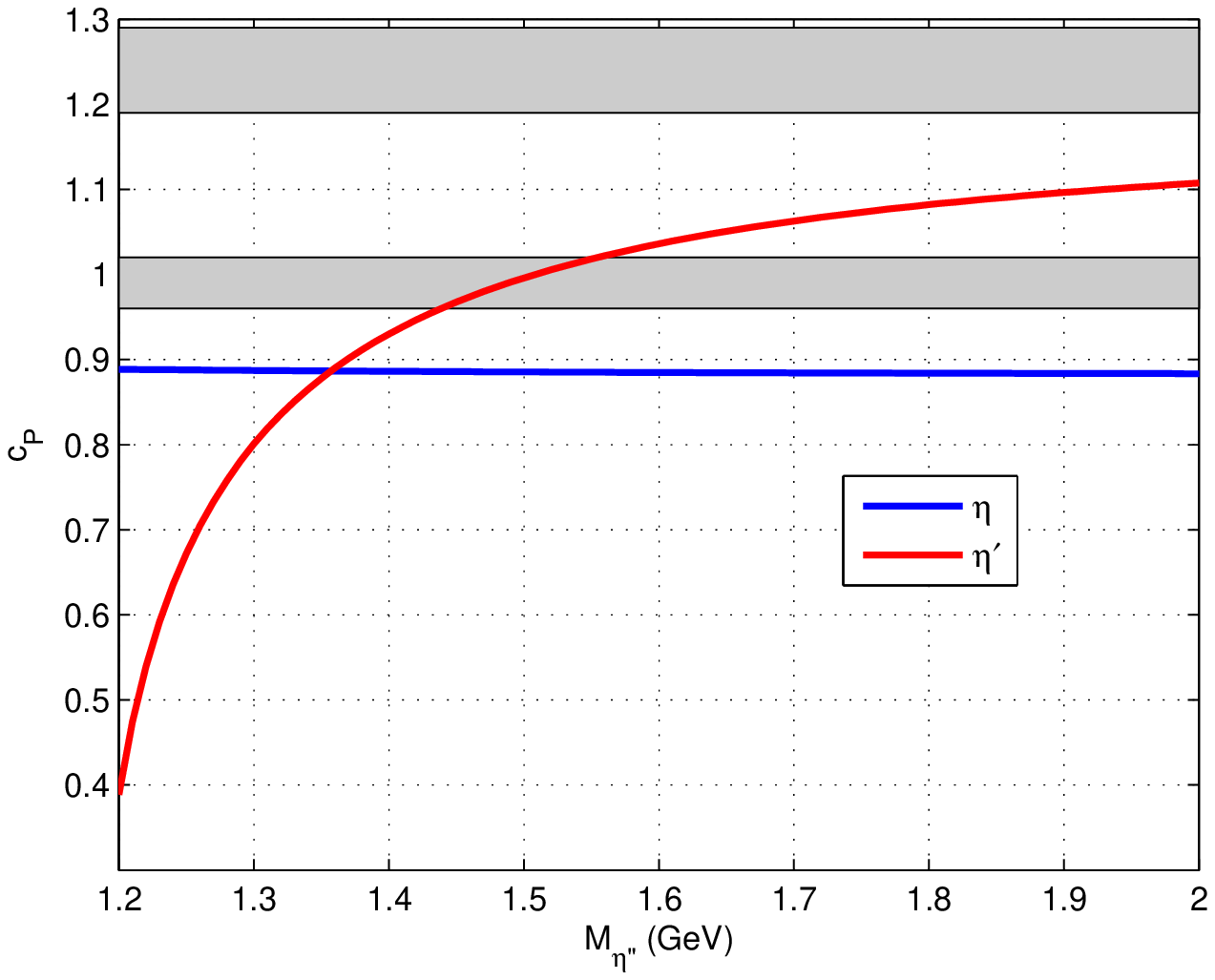}
        \caption{\label{fig:electrans}Electromagnetic transitions and two photons decays.}
\end{figure}

\begin{figure}[htb]
        \includegraphics[width=0.33\linewidth]{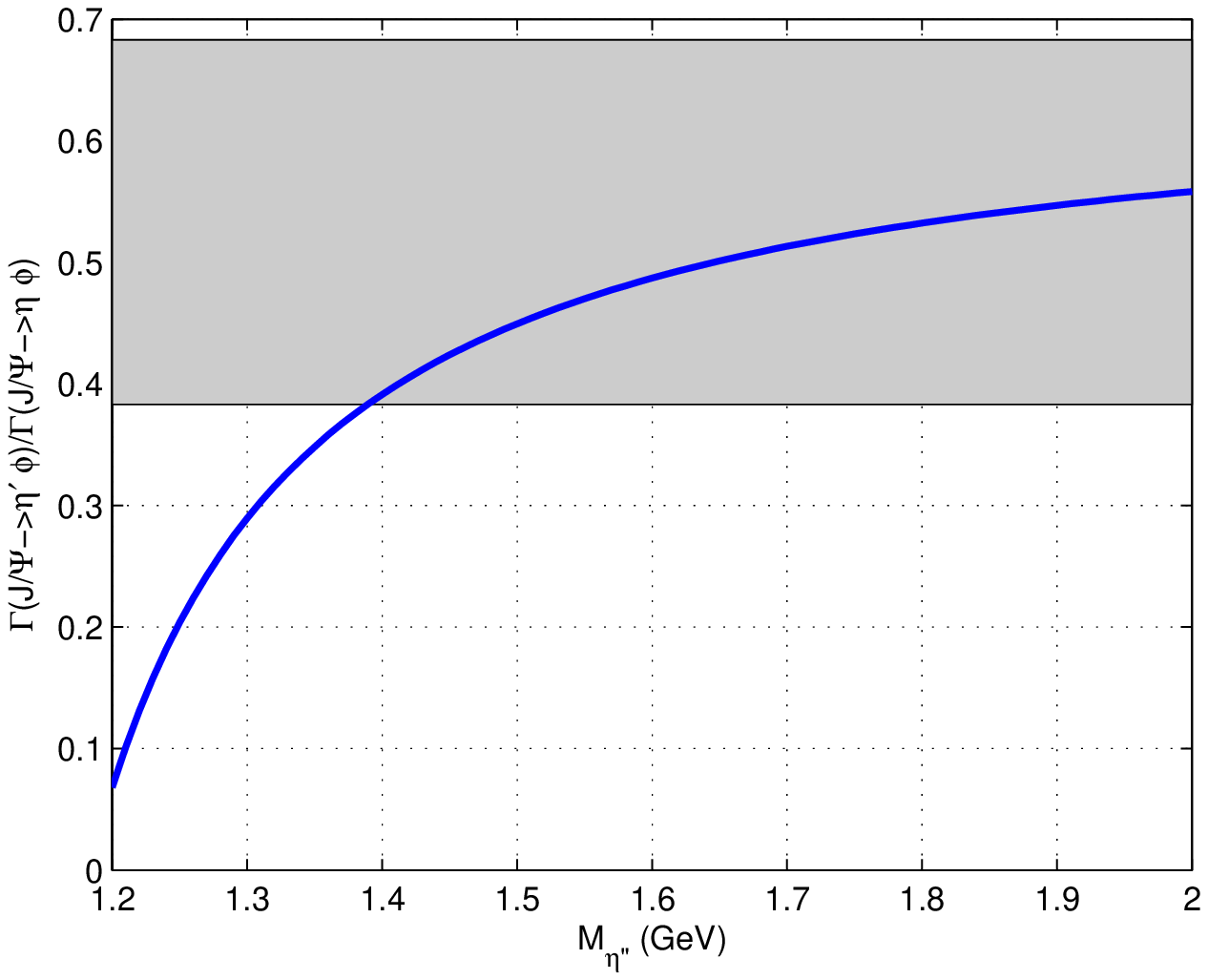}
        \includegraphics[width=0.33\linewidth]{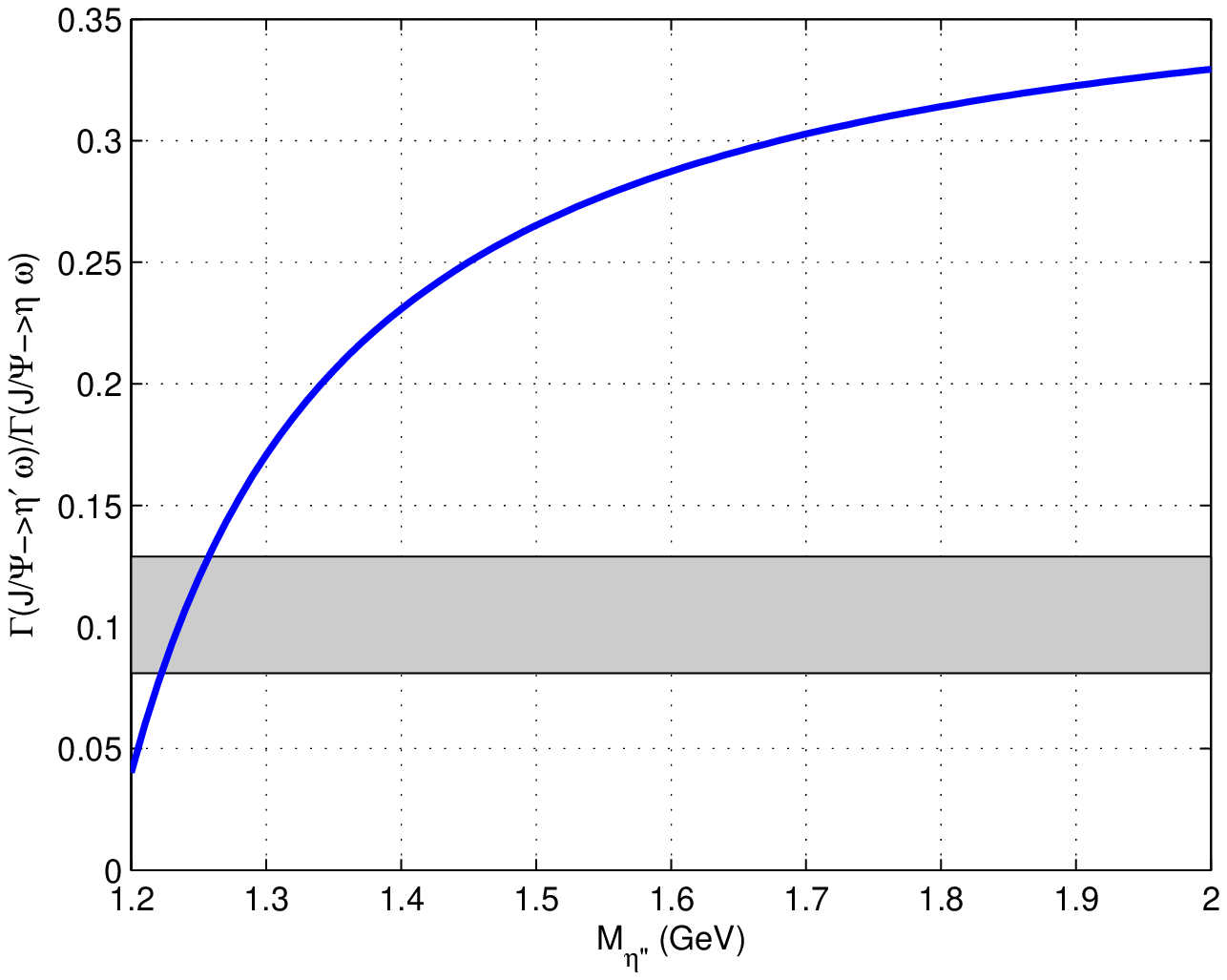}
        \includegraphics[width=0.33\linewidth]{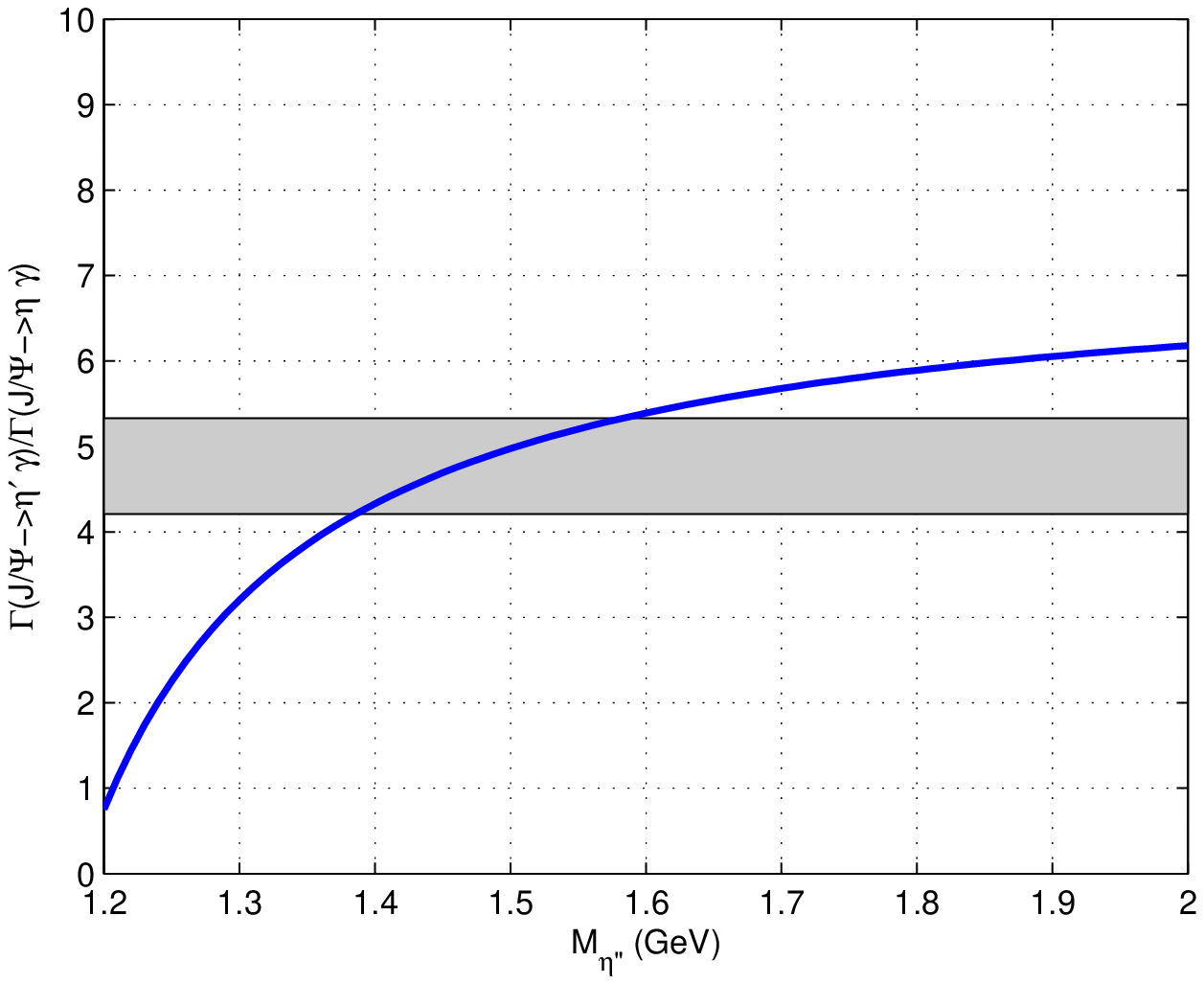}
        \caption{\label{fig:Jpsi}$J/\psi$ decays involving $\eta$ and $\eta'$ mesons.}
\end{figure}

In the large-$N$ approximation, the flavour basis is preferred~\cite{Mathieu:2010ss} and the mass matrix reads in this basis
\begin{equation*}
    {\cal M}_{qsg}^2 = \begin{pmatrix}m_\pi^2 +2\alpha& \sqrt{2}\alpha & \sqrt{2}\beta\\
    \sqrt{2}\alpha & 2m_K^2-m_\pi^2 + \alpha & \beta \\
    \sqrt{2}\beta& \beta & \gamma \end{pmatrix}
\end{equation*}
This matrix can be diagonalized in term the three physical masses~\cite{Mathieu:2009sg}. Adding the leading order interacting Lagrangian for electromagnetic decays and $J/\psi$ decays ($\psi^\alpha$ is the $J/\psi$ field, $Q$ the charge matrix, $V$ the vector meson - the free Lagrangian for vector mesons is understood, $F_{\mu\nu}$ is field strength for the photon and $\alpha=1/137$),
\begin{equation}\label{Lag_gamma}
    {\cal L}_{\gamma} = g_{\gamma} \epsilon_{\alpha\beta\mu\nu}F^{\alpha\beta} \partial^\mu\left\langle Q(V^\nu \pi+\pi V^\nu)\right\rangle + g_{\psi} \epsilon_{\alpha\beta\mu\nu}\partial^\alpha\psi^\beta \partial^\mu\left\langle V^\nu \pi\right\rangle -\frac{N\alpha}{4\pi}F_{\mu\nu}\tilde F^{\mu\nu}\left\langle Q^2 U\right\rangle,
\end{equation}
we can now test our framework on various processes (details have to be presented elsewhere~\cite{DMG}). We have only three free parameters that can be equivalently be the three low energy constant ($\alpha, \beta,\gamma$), the three mixing angles ($\theta,\varphi_G, \varphi$) or, our choice, the three physical masses ($M_{\eta}, M_{\eta'}, M_{\eta''}$). Once one of the set of three parameters is given, branching ratios for various decays follow. Since we perform a leading order analysis, we would like to reproduce the two well-know $\eta$ and $\eta'$ up to 10$\%$. A possible choice for the parameters lying in this range is
\begin{equation}\label{}
    M_{\eta}=530 \text{ MeV} \qquad M_{\eta'}=1030 \text{ MeV}
\end{equation}
The mass of the hypothetical third partner $\eta''$ is left undetermined. Surprisingly, we find an overall agreement for all decays (except the $J/\psi\to\omega\eta(')$ problematic even in the absence of glue) for $M_{\eta''}=1400-1500$ MeV, see Fig.~\ref{fig:electrans} for electromagnetic transitions and Fig.~\ref{fig:Jpsi} for $J/\psi\to PV$ processes.

\begin{figure}[htb]
        \includegraphics[width=0.33\linewidth]{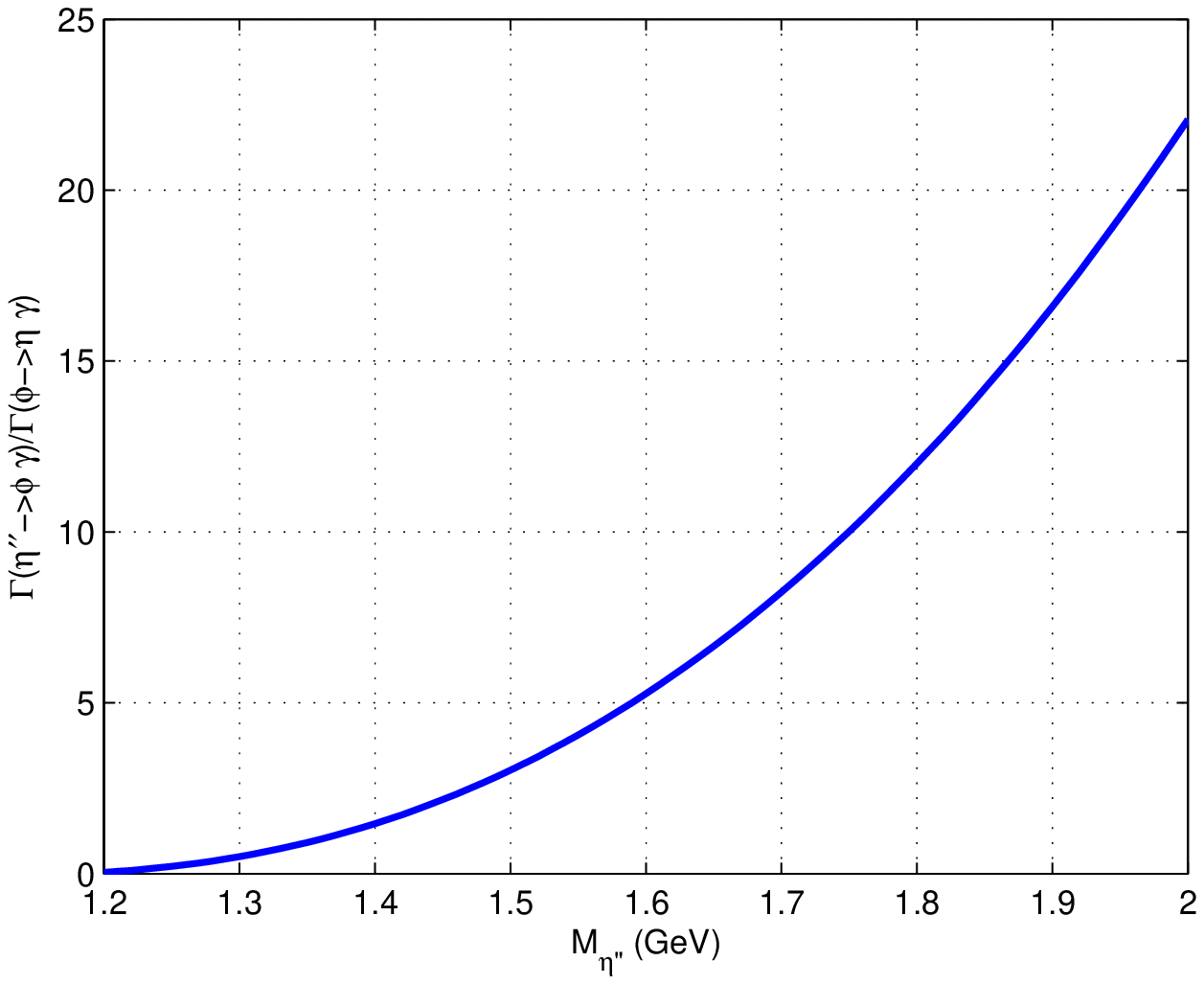}
        \includegraphics[width=0.33\linewidth]{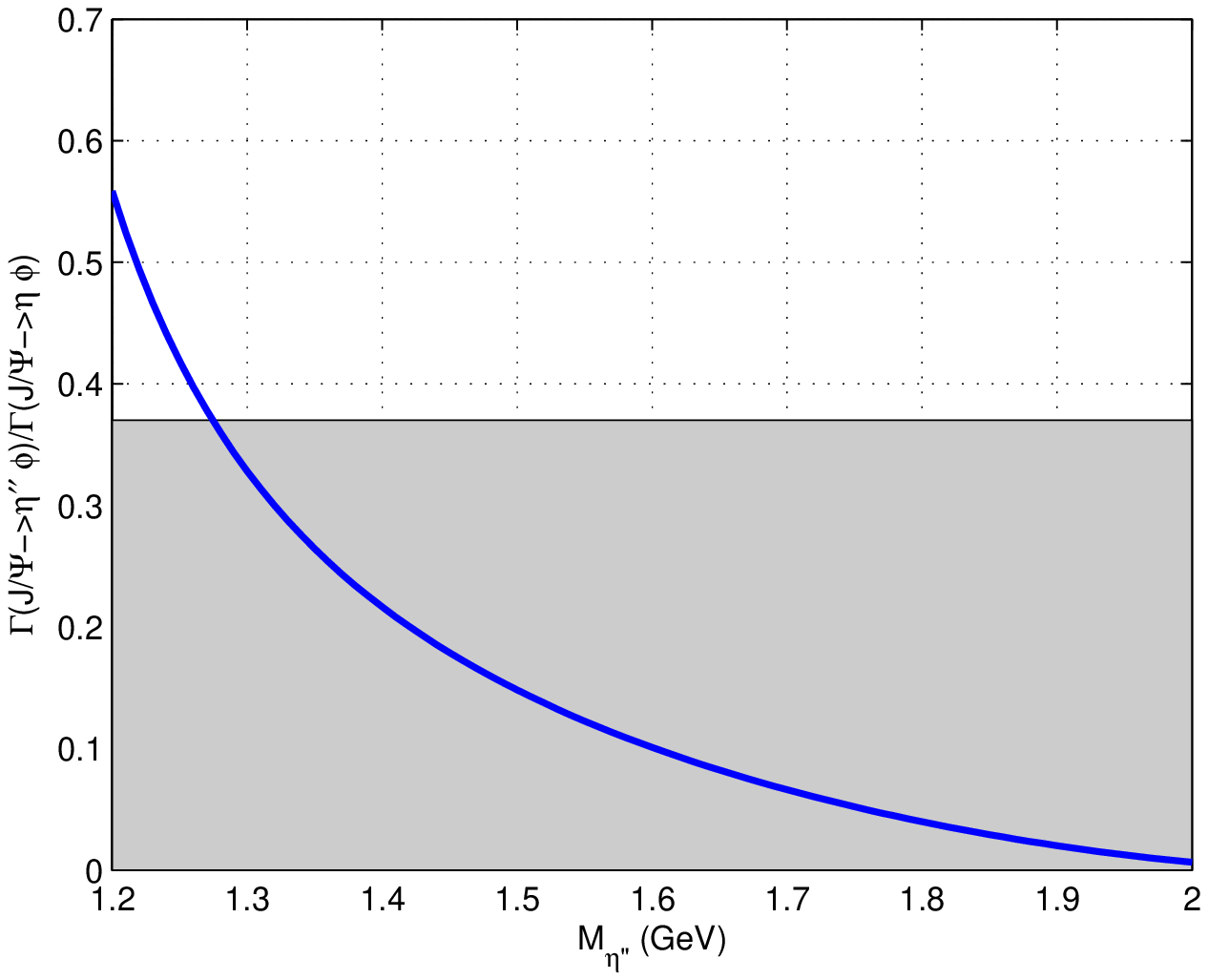}
        \includegraphics[width=0.33\linewidth]{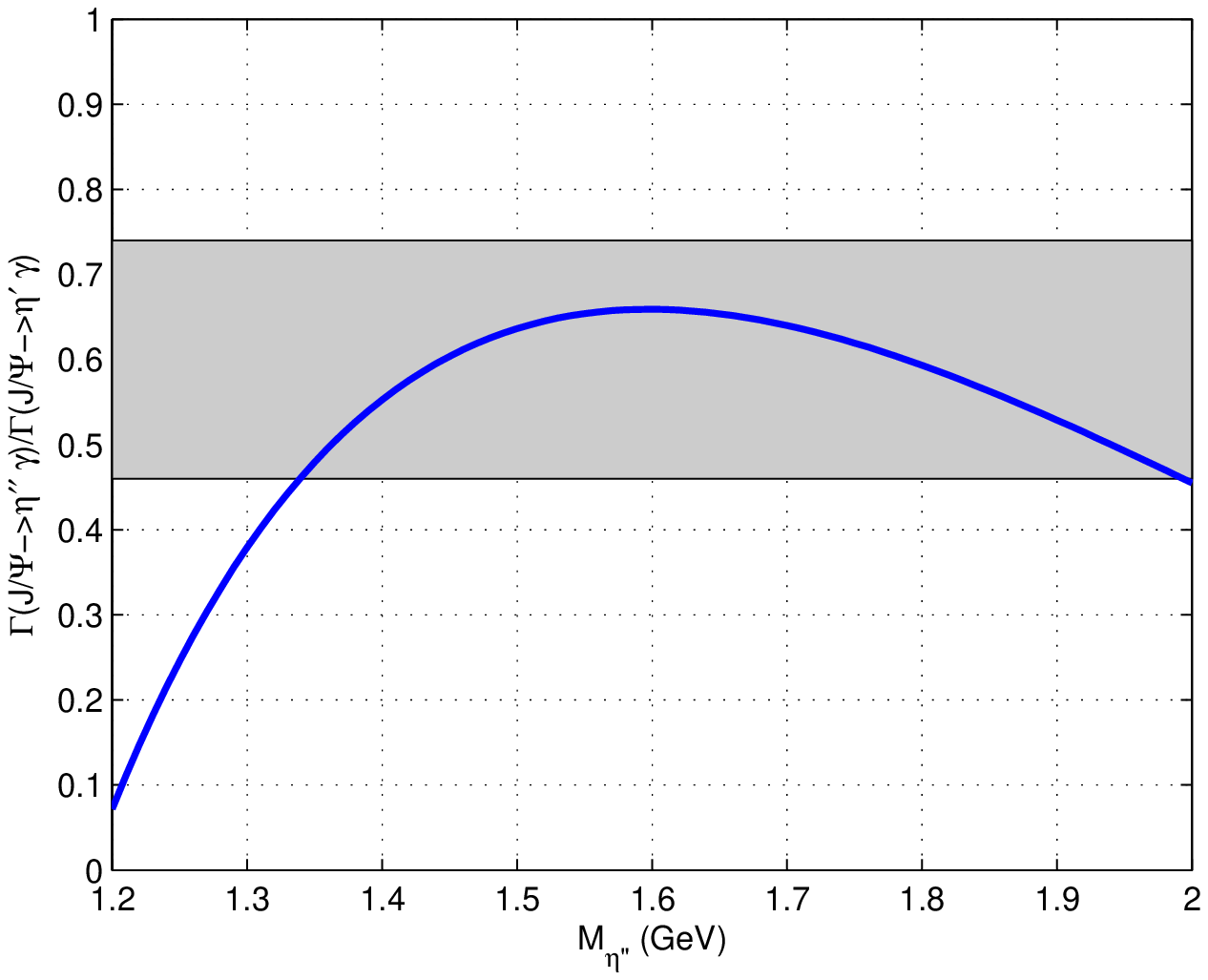}
        \caption{\label{fig:etapp}Decays involving $\eta''$.}
\end{figure}

This constatation encourages us to consider the possibilities that our $\eta''$ is actually the $\eta(1405)$. This possibilities is strengthened by the process $J/\psi\to \eta''\gamma$ showed in Fig.\ref{fig:etapp}. Nevertheless, the lack of data for other processes involving $\eta(1405)$ forbid us to draw definitive conclusion. However, it is still possible to predict branching ratios and hope they will be mesure in the near future. Examples for decays involving $\phi$ and $\eta''$ is also displayed in Fig.~\ref{fig:etapp}.

\section{Conclusion}
In this proceedings, we briefly reviewed the phenomenological evidences for a dynamically generated gluon mass. We explained that although the gluon gains a mass, it behaves like a transverse particle (with only two degrees of freedom). In the pure gauge sector we emphasized on the importance of instanton contributions for (pseudo)scalar glueballs.

The present experimental and theoretical status of the glueball are still ambigus. Although three isoscalar seems to be observed, no definitive conclusion  can be drawn concerning the quark and glue content of those states. In the pseudoscalar sector, however, the situation is a little bit clearer with two well-established states $\eta$ and $\eta'$ and a glueball candidate $\eta(1405)$.

We presented a model based in the chiral Lagrangian to described the $\eta-\eta'-$glue system. Preliminary results favors $\eta(1405)$ to be the glueball partner of $\eta$ and $\eta'$. Predictions are given for various processes involving this $\eta''$. The few data available supports the $\eta(1405)$ interpretation for our $\eta''$ and we hope that future measurements will confirm (or infirm !) our theoretical framework.

\section*{Acknowledgements}
I acknowledge several discussions with the participants during lunches, coffee breaks but also in bars ! This work has been partially funded by the Spanish Ministerio de Ciencia y
Tecnolog\'{\i}a and UE FEDER under contract No. FPA2007-65748-C02-01 and FPA2010-21750-C02-01, by the spanish Consolider Ingenio 2010 Program CPAN (CSD2007-00042) and by the Prometeo
Program (2009/129) of the Generalitat Valenciana. It is also partly funded by
HadronPhisics2, a FP7-Integrating Activities and Infrastructure Program of the
EU under Grant 227431.


\begin{thebibliography}{99}
\bibitem{Cornwall:1981zr}
  J.~M.~Cornwall,
  Phys. Rev. D {\bf 26} (1982) 1453.

\bibitem{PTbook}
  J.M. Cornwall, J.~Papavassiliou, D.~Binosi,\\
  ``The Pinch Technique and Applications to Non-Abelian Gauge Theories''
(Cambridge University Press, Cambridge, 2011).

\bibitem{Aguilar:2008xm}
 A.~C.~Aguilar, D.~Binosi and J.~Papavassiliou,
 Phys.\ Rev.\  D {\bf 78}, 025010 (2008)
 [arXiv:0802.1870 [hep-ph]].

\bibitem{Binosi:2009qm}
  D.~Binosi and J.~Papavassiliou,
  Phys.\ Rept.\  {\bf 479}, 1 (2009)
  [arXiv:0909.2536 [hep-ph]].


\bibitem{Courtoy:2011mf}
  A.~Courtoy, S.~Scopetta, V.~Vento,
  [arXiv:1102.1599 [hep-ph]].

\bibitem{Dokshitzer:1995zt}
  Y.~L.~Dokshitzer, B.~R.~Webber,
  Phys.\ Lett.\  {\bf B352 } (1995)  451-455.

\bibitem{Field:2001iu}
  J.~H.~Field,
  Phys.\ Rev.\  {\bf D66 } (2002)  013013.



\bibitem{Cucchieri:2007md}
  A.~Cucchieri, T.~Mendes,
  PoS {\bf LAT2007 } (2007)  297
\bibitem{Bogolubsky:2007ud}
  I.~L.~Bogolubsky, E.~M.~Ilgenfritz, M.~Muller-Preussker {\it et al.},
  PoS {\bf LAT2007 } (2007)  290
\bibitem{Oliveira:2009nn}
  O.~Oliveira, P.~J.~Silva,
  PoS {\bf QCD-TNT09 } (2009)  033.

\bibitem{Mathieu:2008me}
  V.~Mathieu, N.~Kochelev and V.~Vento,
  Int.\ J.\ Mod.\ Phys.\ E {\bf 18} (2009) 1


\bibitem{Deur:2009tj}
  A.~Deur,
  AIP Conf.\ Proc.\  {\bf 1149 } (2009)  281-284.
  [arXiv:0901.2190 [hep-ph]].




\bibitem{Schwinger:1962tn}
  J.~S.~Schwinger,
  Phys.\ Rev.\  {\bf 125 } (1962)  397-398.
\bibitem{Jackiw:1973tr}
  R.~Jackiw and K.~Johnson,
  Phys.\ Rev.\  D {\bf 8} (1973) 2386 ;
  E.~Eichten and F.~Feinberg,
  Phys.\ Rev.\  D {\bf 10} (1974) 3254 ;
  E.~C.~Poggio, E.~Tomboulis and S.~H.~Tye,
  Phys.\ Rev.\  D {\bf 11} (1975) 2839.



\bibitem{Cornwall:1974hz}
  J.~M.~Cornwall,
  Phys.\ Rev.\  {\bf D10 } (1974)  500.


\bibitem{Mathieu:2009cc}
  V.~Mathieu,
  PoS {\bf QCD-TNT09 } (2009)  024.
  [arXiv:0910.4855 [hep-ph]].
  V.~Mathieu, F.~Buisseret and C.~Semay,
  { Phys.\ Rev.\  D }{\bf 77} (2008) 114022
  [arXiv:0802.0088 [hep-ph]].
  N.~Boulanger, F.~Buisseret, V.~Mathieu and C.~Semay,
  Eur.\ Phys.\ J.\  A {\bf 38}, 317 (2008)




\bibitem{Morningstar:1999rf}
  C.~J.~Morningstar and M.~J.~Peardon,
  { Phys.\ Rev.\  D} {\bf 60}, 034509 (1999)

\bibitem{Jaffe:1985qp}
  R.~L.~Jaffe, K.~Johnson, Z.~Ryzak,
  Annals Phys.\  {\bf 168 } (1986)  344.

\bibitem{Forkel:2003mk}
  H.~Forkel,
  Phys.\ Rev.\  D {\bf 71}, 054008 (2005)

\bibitem{Dudal:2010cd}
  D.~Dudal, M.~S.~Guimaraes, S.~P.~Sorella,
  Phys.\ Rev.\  Lett. {\bf 106}, 062003 (2011)



\bibitem{Crede:2008vw}
  V.~Crede and C.~A.~Meyer,
  Prog.\ Part.\ Nucl.\ Phys.\  {\bf 63}, 74 (2009)

\bibitem{Close:2000yk}
  F.~E.~Close, A.~Kirk,
  Phys.\ Lett.\  {\bf B483 } (2000)  345-352.


\bibitem{Chanowitz:2005du}
  M.~Chanowitz,
  Phys.\ Rev.\ Lett.\  {\bf 95 } (2005)  172001.

\bibitem{Cheng:2006hu}
  H.~-Y.~Cheng, C.~-K.~Chua, K.~-F.~Liu,
  Phys.\ Rev.\  {\bf D74 } (2006)  094005.




\bibitem{Rosenzweig:1981cu}
  C.~Rosenzweig, A.~Salomone and J.~Schechter,
  Phys.\ Rev.\  D {\bf 24}, 2545 (1981).

\bibitem{Mathieu:2009sg}
  V.~Mathieu and V.~Vento,
  Phys.\ Rev.\  D {\bf 81}, 034004 (2010)


\bibitem{Mathieu:2010ss}
  V.~Mathieu, V.~Vento,
  Phys.\ Lett.\  {\bf B688 } (2010)  314-318.
  [arXiv:1003.2119 [hep-ph]].


\bibitem{DMG}
C. Degrande, V. Mathieu and J.-M. Gerard, in preparation










\end{thebibliography}
\end{document}